# Automatic Nested Loop Acceleration on FPGAs Using Soft CGRA Overlay

Cheng Liu, Ho-Cheung Ng and Hayden Kwok-Hay So
Department of Electrical and Electronic Engineering
The University of Hong Kong
Email: {liucheng, hcng, hso}@eee.hku.hk

*Abstract*—Offloading compute intensive nested loops to execute on FPGA accelerators have been demonstrated by numerous researchers as an effective performance enhancement technique across numerous application domains. To construct such accelerators with high design productivity, researchers have increasingly turned to the use of overlay architectures as an intermediate generation target built on top of off-the-shelf FPGAs. However, achieving the desired performance-overhead trade-off remains a major productivity challenge as complex application-specific customizations over a large design space covering multiple architectural parameters are needed.

In this work, an automatic nested loop acceleration framework utilizing a regular soft coarse-grained reconfigurable array (SCGRA) overlay is presented. Given high-level resource constraints, the framework automatically customizes the overlay architectural design parameters, high-level compilation options as well as communication between the accelerator and the host processor for optimized performance specifically to the given application. In our experiments, at a cost of 10 to 20 minutes additional tools run time, the proposed customization process resulted in up to 5 times additional speedup over a baseline accelerator generated by the same framework without customization. Overall, when compared to the equivalent software running on the host ARM processor alone on the Zedboard, the resulting accelerators achieved up to 10 times speedup.

## I. INTRODUCTION

Offloading compute intensive nested loops to FPGA accelerators has been demonstrated by many researchers to be an effective solution for performance enhancement across many application domains [1], [2]. However, the relatively low productivity in developing FPGA-based compute applications remains one of the major obstacles that hinder widespread employment of FPGAs [3]. To address this challenge, a number of researchers have turned to the use of virtual FPGA overlay architectures built on top of the physical FPGA configurable fabric to help with improving design productivity through fast compilation, good design portability and debugging support [4], [5], [6], [7], [8], [9], [10], [11].

Despite the great advantages on design productivity, the additional layer on top of the physical FPGA inevitably introduces performance and resource consumption penalty. An overlay must ensure that the overall FPGA acceleration performance remains competitive. Otherwise, mapping the loop kernels to the overlay based FPGA accelerators will not be as useful. Therefore, the capability to customize the overlay specifically to an application or a domain of application becomes essential to the overlay based FPGA accelerator design. However, navigating through a labyrinth of architectural and compilation parameters to fine-tune an accelerator's performance is a slow and non-trivial process. To require a user to manually explore such vast design space is going to counteract the productivity benefit of the utilizing overlay in the first place.

We have been developing in-house a soft coarse-grained reconfigurable array (SCGRA) overlay based nested loop acceleration framework targeting a hybrid CPU-FPGA system called QuickDough, which allows rapid compilation from C loops to FPGA with a library of pre-built overlay bitstreams [10]. In this work, we mainly focus on automatically customizing the overlay architectural parameters, exploiting loop unrolling and hardware-software communication in combination with buffer sizing specifically to an application with given high-level resource constraints. In particular, by taking advantage of the regularity of the SCGRA overlay, a multitude of design metrics such as performance and hardware consumption can be accurately estimated using analytical models once the overlay scheduling result is available. While the overlay scheduling depends on much less design parameters, the overall customization framework can be dramatically simplified. With both the efficient application-specific customization and rapid compilation, the proposed design framework ensures both high design productivity and high performance of FPGA loop acceleration.

From our experiments, it took the framework 10 to 20 minutes to complete the loop accelerator customization using our proposed two-step approach, which was up to 100 times faster than an exhaustive search through the design space. With customization, the resulting accelerators performed up to 5 times faster than a corresponding baseline accelerator before customization. Overall, when compared to the performance of the benchmark executed on the host ARM processor, the resulting FPGA accelerators achieved up to $10\times$ speedup.

## II. RELATED WORK

Overlay architecture which is a virtual intermediate architecture overlaid on top of off-the-shelf FPGA is increasingly applied as a way to address the productivity challenge.

Various overlays with diverse configuration granularities and flexibility ranging from virtual FPGAs [4], [6], [5], array-of-FUs [7], [8], [11], soft CGRA [9], [10], soft GPU [12], vector processors[13], [14] to configurable processors or multi-core processors [15], [16], [17], [18], [19], [20] have been developed over the years. SCGRA overlay provides unique advantages on compromising hardware implementation and performance for compute intensive nested loops as demonstrated by numerous ASIC CGRAs [21], [22]. Most importantly, it allows both rapid compilation by taking advantage of the overlays' tiling structure [23] and efficient bitstream reuse within the design iterations of an application [10], thus it is particularly promising for high productivity nested loop acceleration.

In addition, customizing the CGRA specifically for an application or a domain of application provides promising performance improvement while saving the hardware resource at the same time as demonstrated in CGRA work targeting ASIC design [24], [25], [26]. While CGRA customization on ASIC is relatively limited due to the tape-out cost, CGRA overlays allow more intensive architectural customization providing just enough hardware to the target application or application domains because of the FPGA's inherent programmability. In [27], Coole and Stitt proposed to provide the overlay with limited flexibility instead of full configurability specifically to a group of design. With this customization, the area overhead was reduced significantly. The authors in [28] developed an SCGRA topology customization method using genetic algorithm and showed the potential benefits of the SCGRA overlay customization. Nevertheless, the rest of the system design parameters were not covered. In [2], the authors formalized the loop acceleration on a regular processing array overlay on FPGA. They focused on the hardware resource constrain, IO bandwidth constrain and the loop parallelism partition while processing architectural design parameters were not included. In order to achieve both high design productivity and high performance with low overhead, a complete nested loop acceleration framework targeting CPU-FPGA system is developed in this work. It supports intensive application-specific customization including the overlay architectural customization, the compilation customization and communication interface customization for optimized performance.

## III. NESTED LOOP ACCELERATOR DESIGN FRAMEWORK

By using a regular SCGRA overlay built on top of the physical FPGA devices, we have developed an automatic nested loop acceleration framework called QuickDough. QuickDough targets hybrid CPU-FPGA computing systems where the FPGA is devoted to accelerating compute intensive loop kernel and CPU handles the rest of the application. Figure 1 depicts an overview of the design framework, highlighting the complementary *accelerator generation* and *accelerator customization* paths.

By design, the steps along the accelerator generation path are short and essential during rapid design iterations. Collectively, they are able to generate FPGA loop accelerators making use of a pre-built bitstream library in the order of seconds [10].

Meanwhile, the focus of this paper is on the accelerator customization path, which is relatively slow but is necessary for improving performance of the resulting accelerators on a per-application basis. These steps automatically tunes the design parameters including overlay architectural parameters, compilation options as well as communication between the FPGA accelerator and host processor specifically to a user application under user constraints such as hardware resource budgets. With the customized design parameters, HDL models of the corresponding SCGRA overlay and their associated drivers are then generated. Afterwards, the drivers will be used by the software compiler while the FPGA accelerator will be implemented and stored in the accelerator library, which can be reused by the fast accelerator generation path in subsequent compilations.

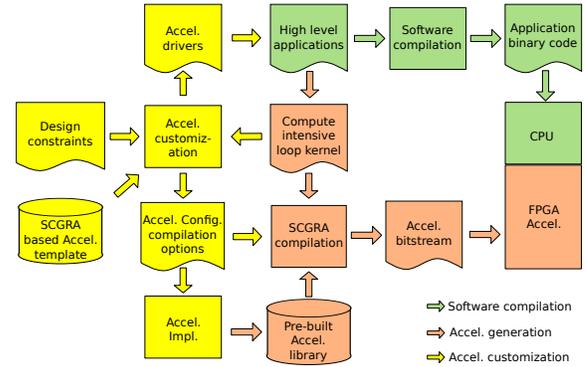

Fig. 1. Automatic nested loop acceleration framework

### A. SCGRA based FPGA accelerator

Figure 2 shows the design of a typical SCGRA overlay based FPGA accelerator. In the accelerator, on-chip memory i.e. IBuf and OBuf are used to buffer the communication data between the host CPU and the accelerator. A controller is also presented in hardware to control the operations of the accelerator as well as memory transfers. The SCGRA, which is the kernel computation fabric, consists of an array of PEs and it achieves the computation task through the distributed control words stored in each PE. The AddrBuf stores all the valid IO buffer accessing addresses of the computation. The current implementation of a PE template is also presented in this figure. At the heart of the PE is an ALU, which is supported by a multi-port data memory and an instruction memory. Data memory stores intermediate data during the computation while instruction memory stores all control words that determines the action of the PE. In addition, a global signal from the AccCtrl block controls the start/stop of all PEs in the array.

### B. Loop execution on the FPGA accelerator

Figure 3 illustrates how the loop is executed on the FPGA accelerator. First of all, data flow graph (DFG) is extracted from the loop and then it is scheduled on to the SCGRA overlay based FPGA accelerator. Depending on how much the loop is unrolled and transformed to DFG, the DFG may be executed repeatedly until the end of the original loop. In addition, data transfers for multiple executions of the same DFG are batched into groups as shown in Figure 3. On the one hand, this technique is used to reduce the number of batching, which further helps to amortize the initial communication cost. On the other hand, it also results in larger on-chip memory overhead. The proposed customization framework can be used to make the right design choices to achieve an optimal design.

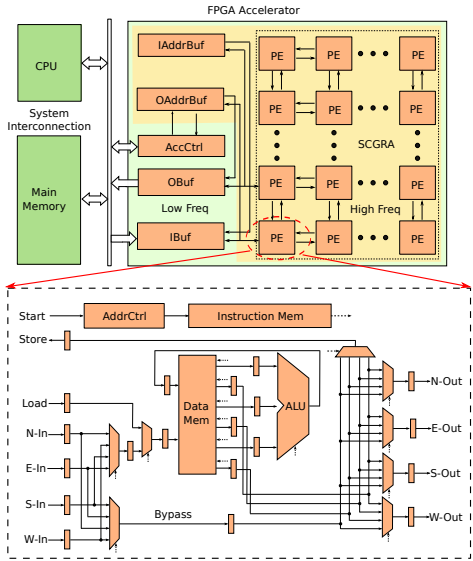

Fig. 2. SCGRA overlay based FPGA accelerator

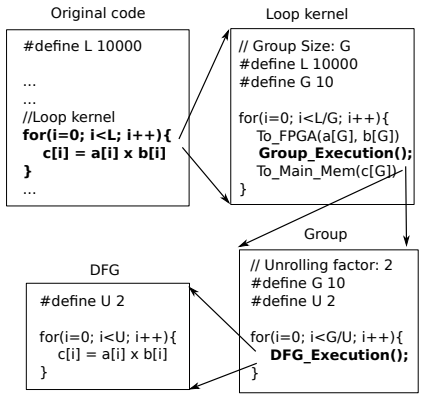

Fig. 3. Loop, group and DFG. The loop will be divided into groups. Each group will be partially unrolled and the unrolled part will be translated to DFG. IO transmission between FPGA and host CPU is performed in the granularity of a group.

### C. SCGRA overlay compilation

With pre-built SCGRA overlay library and customized overlay configuration, the corresponding FPGA accelerator can be generated rapidly, which is also the basis of the high-productivity loop accelerator design framework. Figure 4 presents the detailed SCGRA overlay compilation. With the specified loop unrolling and grouping factor, DFG is generated and scheduled to the SCGRA overlay of the accelerator. After the scheduling, control words are extracted, and they can further be integrated into the pre-built FPGA accelerator bitstream creating the final FPGA loop accelerator bitstream. The compilation process typically completes in a few seconds

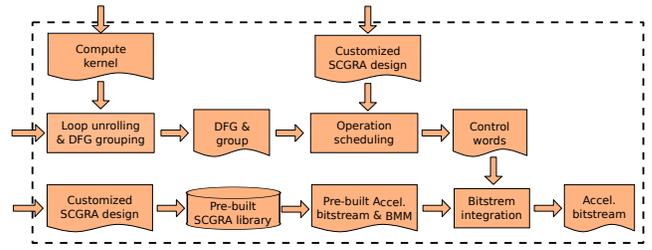

Fig. 4. Rapid SCGRA overlay compilation

as illustrated in [10] which is particularly important during early application development phases.

## IV. SCGRA OVERLAY BASED FPGA ACCELERATOR CUSTOMIZATION

Application-specific customization provides unique opportunity to reduce the resource consumption and improve performance of the resulting accelerators. However, taking the system as a black box and exhaustively searching all the possible configurations can be inefficient and slow. In this work, by taking advantage of the regularity of the SCGRA overlay based FPGA accelerator, we can reduce the complex customization problem to a much simpler sub design space exploration (DSE) together with a simplified search problem. With the customization, optimized application-specific nested loop accelerator can be produced efficiently.

### A. Customization problem formulation

In this section, we will formalize the customization problem of the nested loop acceleration on an SCGRA overlay based FPGA accelerator. Various design constraints including energy consumption and hardware resource consumption can be used while hardware resource consumption is taken as an example here.

TABLE I
DESIGN PARAMETERS OF NESTED LOOP ACCELERATION [1]

| Design Parameters | | Denotation |
|---|---|---|
| Nested Loop Compilation | Loop Unrolling Factor | $\boldsymbol{u} = (u_0, u_1, ...)$ |
| | Grouping Factor | $\boldsymbol{g} = (g_0, g_1, ...)$ |
| Overlay Configuration | SCGRA Topology | 2D Torus, fixed |
| | SCGRA Size | $r \times c$ |
| | Data Width | $W_0$ |
| | Data Mem | $D_0 \times W_0$ |
| | Input Buffer | $D_1 \times W_0$ |
| | Output Buffer | $D_2 \times W_0$ |
| | Instruction Mem | $D_3 \times W_1$ |
| | Input Address Buffer | $D_4 \times W_2$ |
| | Output Address Buffer | $D_5 \times W_3$ |
| | Operation Set | fixed |
| | Implementation Frequency | $f$, fixed |
| | Pipeline Depth | fixed |

Suppose $\boldsymbol{\Psi}$ represents the overall nested loop acceleration design space. $C \in \boldsymbol{\Psi}$ represents a possible configuration in the design space and it includes a number of design parameters as listed in Table I. Assume that the loop to be accelerated has $n$ nested levels and loop count can be denoted as

[1] The parameters are all customizable in the proposed design framework except the ones that are clearly identified as fixed.

$l = (l_1, l_2, ..., l_n)$. $R = (R_1, R_2, R_3, R_4)$ stands for the FPGA resource (i.e. BRAM, DSP, LUT and FF) that are available on a target FPGA and $ResConsumption(C, i)$ denotes the four different types of FPGA resource consumption. $In(g)$ and $Out(g)$ stand for the amount of input and output of a group. Similarly, $In(u)$ and $Out(u)$ stand for the amount of input and output of a DFG. $DFGCompuTime(C)$ represents the number of cycles needed to complete the DFG computation. $\alpha_i$ and $\beta_i$ are constant coefficients depending on target platform where $i = (1, 2, ...)$. With these denotations, the customization problem targeting minimum run time can be formulated as follows:

Minimize
$$RunTime(C) = CompuTime(C) + CommuTime(C) \quad (1)$$

subject to
$$\begin{aligned} ResConsumption(C, i) &\leq R_i, i = 1, 2, 3, 4 \\ In(g) &\leq D_1 \\ Out(g) &\leq D_2 \\ DFGCompuTime(C) &\leq D_3 \\ \prod_{i=1}^{n} \frac{g_i}{u_i} \times In(u) &\leq D_4 \\ \prod_{i=1}^{n} \frac{g_i}{u_i} \times Out(u) &\leq D_5 \end{aligned} \quad (2)$$

$RunTime(C)$ represents the number of cycles needed to compute the loop on the CPU-FPGA system. It consists of both the time consumed for computing on FPGA and communication between FPGA and host CPU, and it can be calculated using Equation 1.

Since the unrolled part of the loop will be translated to DFG and then scheduled to the SCGRA overlay. Thus the DFG computation time is essentially a function of $\mathbf{u}$, $r$ and $c$, and it can also be denoted by $DFGCompuTime(\mathbf{u}, r, c)$. The nested loop is computed by repeating the same DFG execution, and the nested loop computation can be calculated using Equation 3.

$$CompuTime(C) = \prod_{i=1}^{n} \frac{l_i}{u_i} \times DFGCompuTime(\mathbf{u}, r, c) \quad (3)$$

DMA is typically used for the bulk data transmission. Communication cost per data can be modeled with a piecewise linear function and thus DMA latency can be calculated using $DMA(x)$ where $x$ represents the amount of DMA transmission. The communication time of the whole nested loop can be calculated by Equation 4.

$$CommuTime(C) = \prod_{i=1}^{n} \frac{l_i}{g_i} \times (DMA(In(\mathbf{g})) + DMA(Out(\mathbf{g}))) \quad (4)$$

Hardware resource on FPGA mainly includes DSP, LUT, FF and BRAM (block RAM). LUT, FF and DSP consumption can be roughly estimated with a linear function of SCGRA size and can be calculated using Equation 5. BRAM consumption $ResConsumption(C, 1)$ which is slightly different from LUT, FF and DSP consumption can be calculated precisely based on the memory block configurations.

$$ResConsumption(C, i) = \alpha_i \times r \times c + \beta_i, (i = 2, 3, 4) \quad (5)$$

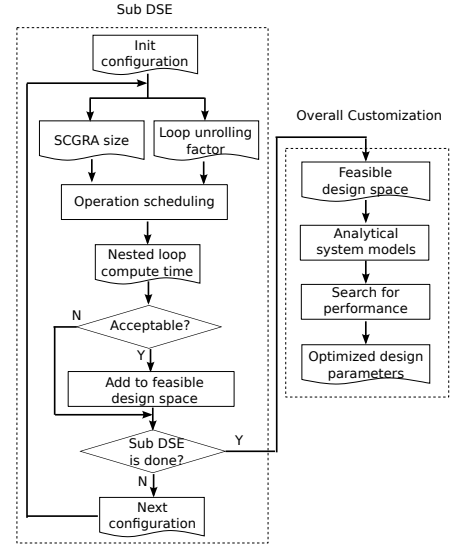

Fig. 5. System customization framework.

### B. Customization framework

Figure 5 illustrates the overview of the customization framework. It can be roughly divided into two parts. In the first part, a sub DSE targeting loop execution time is performed and the feasible design space can be obtained. Since loop execution time is determined by the operation scheduling which simply depends on the loop unrolling factor and SCGRA size, the sub DSE is much simpler compared to the overall system DSE which includes more than 10 design parameters. In the second part, each configuration in the feasible design space will be evaluated. Instead of using simulation based methods, analytical models are employed to estimate the accelerator metrics such as performance and hardware resource consumption. These analytical models are accurate because of the regularity of the SCGRA overlay. Even though the feasible design space is still large, it is fast to evaluate all the configurations in it. After the evaluation process, customization for best performance becomes trivial and the customized design parameters can be obtained immediately.

Suppose $\Phi$ denotes the feasible design space. $\epsilon$ indicates the percentage of the performance benefit obtained by the increase of loop unrolling or SCGRA size. It is a user defined threshold and must be small enough to prune the configurations that are inappropriate. The configurations in $\Phi$ must satisfy Equation 6 and Equation 7.

$$\begin{aligned} &\forall C = (..., \mathbf{u}, r, c, ...) \in \Phi, C' = (..., \mathbf{u'}, r', c', ...) \in \Phi, \\ &(r + 1 == r' \text{ and } c == c') \text{ or } (r == r' \text{ and } c + 1 == c') : \\ &\frac{CompuTime(C) - CompuTime(C')}{CompuTime(C)} > \epsilon \end{aligned} \quad (6)$$

$$\begin{aligned} &\forall C = (..., \mathbf{u}, r, c, ...) \in \Phi, C' = (..., \mathbf{u'}, r, c, ...) \in \Phi, \\ &\mathbf{u} \text{ and } \mathbf{u'} \text{ are consecutive unrolling factors} : \\ &\frac{CompuTime(C) - CompuTime(C')}{CompuTime(C)} > \epsilon \end{aligned} \quad (7)$$

Each feasible configuration $C \in \Phi$ must have gone through the scheduling and thus the corresponding scheduling result is

16

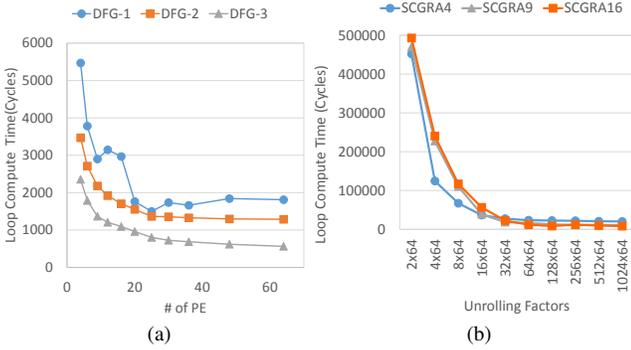

(a)          (b)

Fig. 6. The design parameters typically have monotonic influence on the loop computation time and the computation time benefit degrades with the increase of the design parameter. (a) SCGRA Size, the SCGRA topology used are torus with $2\times2$, $3\times2$, $3\times3$, ... while DFG-1, DFG-2 and DFG-3 are DFGs extracted from matrix-matrix multiplication, fir and Kmean respectively. (b) Unrolling Factor, the loop used is a 63-tap Fir with 1024 input.

known. Consequently, the computation time of the loop kernel and minimum instruction memory depth are available as well. Then we can further evaluate the performance of each feasible configuration using the models built in previous section and obtain the optimized configuration through a simple search.

In addition, a series of experiments on Zedboard as shown in Figure 6 demonstrate that SCGRA size and unrolling factor present a clear monotonic influence on the loop compute time. The performance benefit of loop unrolling and increase of SCGRA size drops gradually. This observation further helps to simplify the sub DSE with a simple branch and bound algorithm.

## V. EXPERIMENTS AND RESULTS

In the experiments, we measured the time needed to customize the loop accelerators and compared the performance of the resulting accelerators to that of an hard ARM processor.

### A. Experiment setup

The customization runtime was obtained using a computer with Intel(R) Core(TM) i5-3230M CPU and 8GB RAM. Zedboard which has an ARM processor and an FPGA was used as the computation system. PlanAhead 14.7 was used for the SCGRA overlay based design. The customized overlay implementations on Zedboard run at 250MHz. To perform the customization, $\epsilon$ is set to be 0.05 and all the resource on Zedboard is set to be the resource constraint. Software runtime is obtained from ARM processor of Zedboard.

In this work, we take four applications including Matrix Multiplication (MM), FIR, Kmean(KM) and Sobel Edge Detector (SE) as our benchmark. The configurations of the benchmark are detailed in Table II.

### B. Customization time

Figure 7 shows the customization time of both the proposed two step (TS) customization and an exhaustive search based customization (ES). TS typically completes the customization in 10 minutes to 20 minutes and it is around 100x faster than the ES on average. In particular, ES is extremely slow on MM which has three levels of loop with relatively large loop count and thus larger design space. Though TS also needs longer time to complete the customization, it skips most of the unfeasible configurations and the runtime is less sensitive to the size of the design space.

TABLE II
BENCHMARK CONFIGURATIONS

| Benchmark | Parameters | Loop Structure |
|---|---|---|
| MM | Matrix Size(100) | $100 \times 100 \times 100$ |
| FIR | # of Input (10000) <br> # of Taps+1 (50) | $10000 \times 50$ |
| SE | # of Vertical Pixels (128) <br> # of Horizontal Pixels (128) | $128 \times 128 \times 3 \times 3$ |
| KM | # of Nodes(5000) <br> # of Centroids(4) <br> # of Dimensions(2) | $5000 \times 4 \times 2$ |

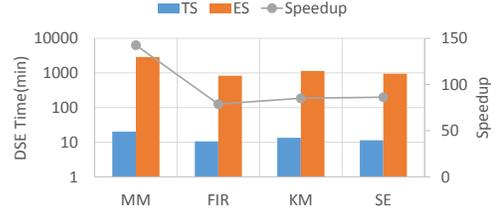

Fig. 7. Benchmark customization time using both TS and ES

TABLE III
ACCELERATOR CONFIGURATIONS [2]

| | | |
|---|---|---|
| MM | Base | $(1 \times 2 \times 100, 4 \times 2 \times 100, 5 \times 5, 1k, 2k)$ |
| | TS | $(1 \times 5 \times 100, 50 \times 5 \times 100, 4 \times 4, 1k, 8k)$ |
| | ES | $(1 \times 5 \times 100, 25 \times 5 \times 100, 5 \times 4, 1k, 8k)$ |
| FIR | Base | $(10 \times 50, 100 \times 50, 3 \times 3, 1k, 2k)$ |
| | TS | $(50 \times 50, 2000 \times 50, 4 \times 4, 1k, 4k)$ |
| | ES | $(100 \times 50, 5000 \times 50, 5 \times 4, 1k, 8k)$ |
| SE | Base | $(4 \times 4 \times 3 \times 3, 128 \times 128 \times 3 \times 3, 3 \times 2, 1k, 8k)$ |
| | TS | $(16 \times 16 \times 3 \times 3, 128 \times 128 \times 3 \times 3, 4 \times 4, 1k, 4k)$ |
| | ES | $(16 \times 16 \times 3 \times 3, 128 \times 128 \times 3 \times 3, 5 \times 4, 1.5k, 4k)$ |
| KM | Base | $(25 \times 4 \times 2, 2500 \times 4 \times 2, 4 \times 3, 1k, 8k)$ |
| | TS | $(125 \times 4 \times 2, 625 \times 4 \times 2, 5 \times 5, 1k, 2k)$ |
| | ES | $(125 \times 4 \times 2, 625 \times 4 \times 2, 5 \times 5, 1k, 2k)$ |

### C. Customized accelerator performance

In order to demonstrate the quality of proposed framework, we compared the performance of the accelerators with a random configuration as well as customized configurations obtained using both TS and ES. The detailed configurations of the accelerators are listed in Table III. The performance comparison is shown in Figure 8. It can be found that the customized accelerators obtained using TS achieve quite close performance to the ones customized through ES. Particularly, the customized accelerator achieves up to 10X speedup over the ARM processor on the benchmark. For FIR, SE and KM, the speedup is promising. MM has relatively low compute-IO rate and the single input and output between the on-chip buffer and the SCGRA overlay limits the performance of the accelerator. This problem can hopefully be alleviated by appropriate on-chip buffer partition, which will be supported in the proposed framework in future.

[2] The configurations include loop unrolling factor, grouping factor, SCGRA array size, instruction memory depth and IO buffer depth

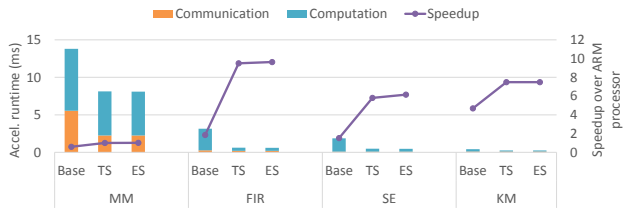

Fig. 8. Customized FPGA loop accelerator performance

## VI. CONCLUSION

In this work, we have presented an automatic nested loop acceleration framework that is based on a soft coarse-grained reconfigurable array overlay. We have demonstrated that by taking advantage of the regularity of the overlay, intensive system customization specific to the given user application can be performed efficiently, resulting in up to 5 times performance improvement over solutions without customization at the cost of 10 to 20 minutes additional tools run time. Overall, the framework is able to generate accelerators that achieve up to 10 times speed up over software running on the host processor, resulting in a high design productivity experience for software programmers.

## ACKNOWLEDGMENT

This work was supported in part by the Research Grants Council of Hong Kong project ECS 720012E and the Croucher Innovation Award 2013.